\documentstyle[twocolumn,prc,aps,epsf]{revtex}

\newcommand{\be}{\begin{eqnarray}}
\newcommand{\ee}{\end{eqnarray}}
\begin{document}
\twocolumn[\hsize\textwidth\columnwidth\hsize\csname @twocolumnfalse\endcsname

\vskip 0.5cm
\title{Comment on ``On the Origin of the OZI Rule in QCD", 
by N.~Isgur and H.~B.~Thacker}
\author{T.~Sch{\"a}fer$^{1,2}$ and E.~Shuryak$^1$}
\address{$^1$Department of Physics, SUNY Stony Brook, 
Stony Brook, NY 11794.\\
$^2$RIKEN-BNL Research Center, Brookhaven National Laboratory,
Upton, NY 11973.}
\maketitle

\vspace{0.5cm}

\begin{abstract}
 We comment on the recent paper (hep-lat/0005006) by Isgur 
and Thacker on the origin of the OZI rule in QCD. We show that 
instantons explain the sign and magnitude of the observed 
OZI-violating amplitude in all mesonic channels, not just in 
the $\eta'$ channel. We comment on the role of instantons in 
hadronic spectroscopy and the relation between instantons and 
the large $N_c$ limit of QCD.
\end{abstract}
\vspace{0.5cm}
]

\begin{narrowtext}   
\newpage 

\section{Introduction}

 In their recent paper \cite{IT}, Isgur and Thacker discuss an issue 
of paramount importance for mesonic spectroscopy, the nature of the 
large OZI-violating amplitude observed in the pseudoscalar nonet. 
The $\pi-\eta'$ splitting is the largest mass splitting among light 
mesons, and understanding its physical origin is clearly a key issue.

 In order to clarify the physical origin of the large $\eta'$ mass
Isgur and Thacker studied its analogue in the scalar $0^{++}$,
vector $1^{--}$, and axialvector $1^{++}$ channels. They observe 
that the OZI-violating amplitude $A^{0^{++}}$ in the scalar channel 
is as large as the one in the pseudoscalar channel. In the first 
version of their paper, Isgur and Thacker claimed that the signs of 
$A^{0^{++}}$ and $A^{0^{-+}}$ are the {\em same}. After being alerted 
to an error in their work by us and other, the second version concludes 
that the two amplitudes are {\em opposite} in sign. As we discuss below,
this sign is crucial for phenomenology, and it is indeed what 
instantons predict it to be. Nevertheless, Isgur and Thacker
still conclude that {\em ``... our result favors the large $N_c$ 
and not the instanton interpretation of the solution of the $\eta'$ 
mass problem"}. To us, this appears to be a significant misunderstanding, 
and we would like to clarify the issue in this comment.  
  
 We emphasize that the issue of the sign of the $A^{0^{++}}$ amplitude
is also connected with an old misunderstanding concerning the scalar 
isoscalar (sigma) meson channel. Because the sigma resonance around 
600 MeV is so broad and has not always been included in the Particle
Data Table, it is sometimes assumed that the lightest state
in this channel is located around 1.5 GeV or higher, and that the 
interaction must therefore be strongly repulsive. But we know that,
fundamentally, the interaction in the $0^{++}$ channel must be very 
attractive, how else could chiral symmetry breaking take place? 
Using models of QCD it is hard to see how, after the vacuum is 
rearranged and chiral symmetry breaking has taken place, the 
$\sigma$ state could be pushed up to a mass much beyond 600 MeV. 
One might argue that there is a an attractive interaction in the 
$0^{++}$ channel, but that it respects the OZI rule. But in this 
case one is faced with the problem that no low mass strength is
seen in the isovector channel, so the $I\!=\!1$ $0^{++}$ channel is 
indeed repulsive. This is clearly seen in lattice calculations 
of the scalar isovector correlation function. 

  We would also like to discuss a somewhat secondary point. Isgur 
and Thacker emphasize that their results are in agreement with 
$1/N_c$ arguments, even though the $1/N_c$ expansion makes no 
prediction concerning the sign of the amplitude. They also 
emphasize that $1/N_c$ arguments are generally incompatible 
with instantons. This is an issue on which there is a lot of
confusion in the literature, and we shall comment on it below. 

  A general issue worth discussing in this note is the question
of ``conspiracies'' among different hadronic amplitudes, which 
Isgur and Thacker mention in connection with the OZI rule in
the vector channel. On this point we completely agree with 
Isgur and Thacker. We discuss other examples of similar 
conspiracies that have appeared in the study of hadronic
correlation functions. Despite our criticism, we consider 
the paper by Isgur and Thacker to be a positive step. The
problems related to OZI-violating amplitudes, chiral symmetry 
breaking, the $U(1)_A$ anomaly, and their relation to the 
foundations of the constituent quark model are rarely discussed 
in the literature on hadronic spectroscopy.

\section{The sign of the OZI-violating amplitude: Scalars vs pseudoscalars}

  Let us first recall the setting. Isgur and Thacker consider QCD with
two flavors and introduce $4\times 4$ mass matrices in the basis of 
$u\bar d, d\bar u,u\bar u, d \bar d$ states of the form 
\begin{equation}
\bf T =
\left[ \matrix{S & 0 &  0  &  0  \cr
               0 & S &  0  &  0  \cr
               0 & 0 & S+A &  A  \cr
               0 & 0 &  A  & S+A \cr} \right] ~~.
\end{equation}
Here, we ignore the effects of quark masses, and the eigenstates have 
to fall into isospin multiplets. It is then sufficient to consider
the lower $2\times 2$ block of the mass matrix
\begin{equation}
\bf T \sim \left[ \matrix{
   D & A  \cr
   A & D  \cr  } \right] ~~.
\end{equation}
where $D=S+A$ in the notation of Isgur and Thacker. So there are two 
amplitudes, $D,A$, and two different eigenstates in every $J^{PC}$ channel. 
In the pseudoscalar channel the physical states are the $\eta'\sim (\bar{u} 
i\gamma_5 u+\bar{d}i\gamma_5 d)$ and $\pi^0\sim(\bar{u}i\gamma_5 u-\bar{d} 
i\gamma_5 d)$, with masses (or mass squared) $D-A$ and $D+A$. We emphasize 
that both $S$ and $A$ depend on the quantum numbers of the current. In the 
case $\Gamma=i\gamma_5$ there is no question about the sign of $A$: we want 
$m_{\eta'}>m_\pi$ and thus $A>0$. 

 Similarly, there are two independent scalar channels, traditionally 
called  $\sigma \sim (\bar{u}u + \bar{d}d)$ and $\delta^0 \sim(\bar{u}u 
- \bar{d}d)$ ($f_0$ and $a_0$ in modern notation). The issue at hand is 
the sign of the corresponding amplitude $A^{0++}$.

 Before we go into phenomenology, let us explain the instanton and
perturbative QCD predictions. As discovered in the classical paper 
by t'Hooft \cite{tHooft} the effect of instantons on fermionic 
correlation functions can be summarized in terms of an effective 
interaction
\begin{equation} 
L = G\left[
  (\bar{\psi}\tau_a\psi)^2 - (\bar{\psi}\psi)^2 
- (\bar{\psi}i\gamma_5\tau_a)^2 + (\bar{\psi}i\gamma_5)^2 \right],
\end{equation}
where $G$ is an effective coupling (that depends on the instanton
amplitude) and $\tau$ is an isospin matrix. We can directly read
off the interaction in the channel characterized by the current 
$\bar\psi\Gamma\psi$. The interaction is attractive in the pion 
$\bar\psi i\gamma_5\vec{\tau}\psi$ and sigma $\bar\psi\psi$
channels, and repulsive in the eta prime $\bar\psi i\gamma_5\psi$
and delta $\bar\psi\vec{\tau}\psi$ channels. So, to first order, 
the instanton-induced interaction corresponds to $A^{0-+}=-A^{0++}$. 

\begin{figure}
\vspace*{-0.6cm}
\begin{center}
\leavevmode
\epsfxsize=7cm
\epsffile{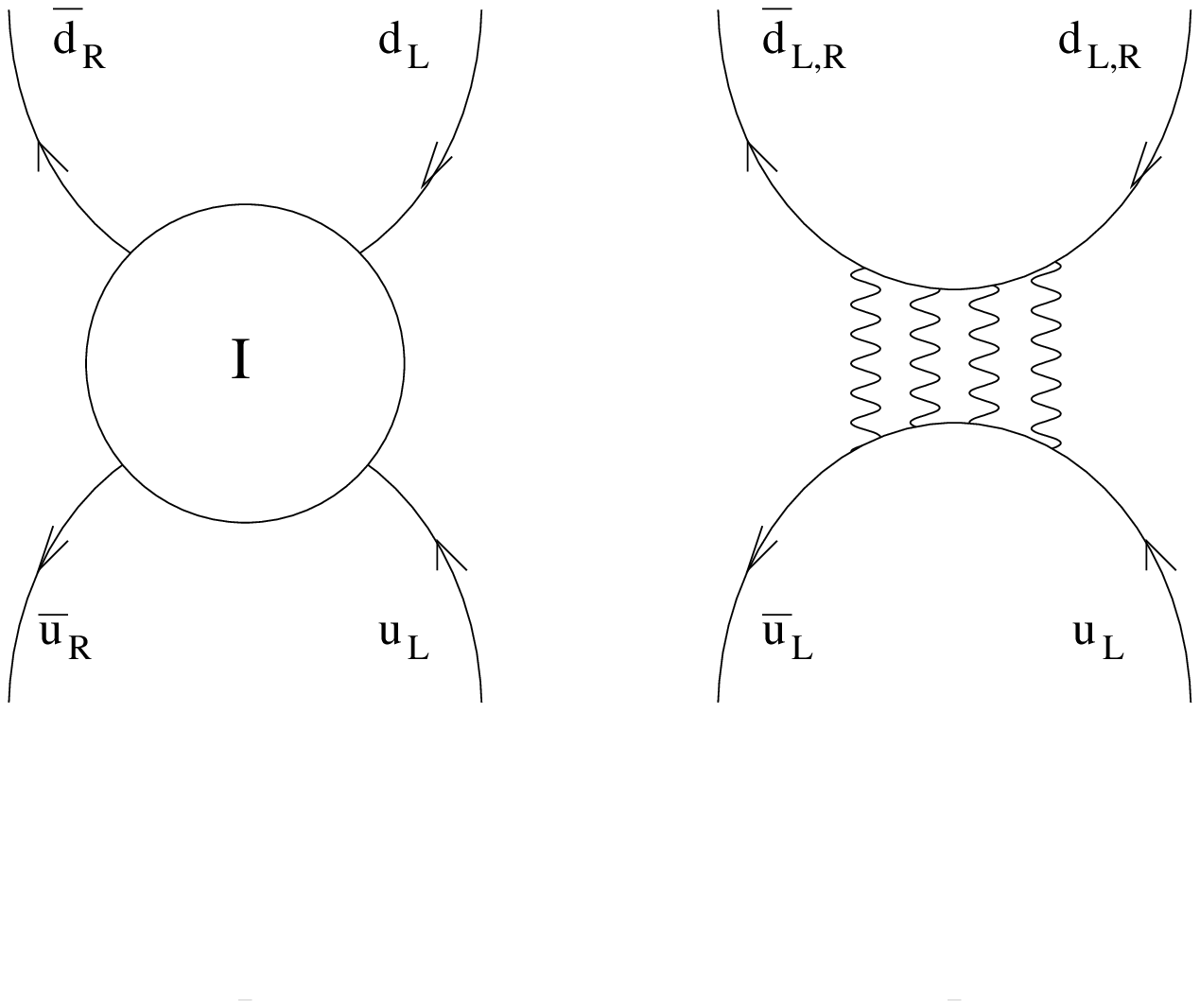}
\vspace*{-1.4cm}
\end{center}

\caption{Instanton (left panel) and perturbative (right panel)
contribution to OZI-violating correlation functions.}
\end{figure}

It is worth repeating why this is so \cite{SS}. The instanton 
interaction corresponds to the contribution of fermionic zero
modes to the quark propagator. Since there is exactly one zero
mode for every flavor, there are no diagonal $(\bar{u}u)(\bar{u}u)$ 
or $(\bar{d}d)(\bar{d}d)$ interactions. Second, since the fermion
zero modes for quarks and anti-quarks have opposite chirality,
the interaction is also off-diagonal in the basis spanned by 
right and left-handed fermions $q_R,\,q_L$ 
\begin{equation}
\bf T \sim\left[ \matrix{
 0 & A \cr
 A & 0 \cr} \right] ~~.
\end{equation}
This means that the instanton-induced amplitude follows very 
simple rules: (i) the sign flips in going from scalar to pseudoscalar, 
(ii) the sign flips in going from $I=0$ to $I=1$ states, (iii) to leading
order there is no interaction in vector channels. This means that the 
OZI-violating amplitude in the vector channels receives no direct instanton 
contribution and is expected to be small. To summarize, we have the 
following prediction for the signs of the instanton contribution
\begin{equation}
\bf 
\left[ \matrix{\eta'  & + \cr
\pi & - \cr
\sigma & - \cr
\delta & + \cr
} \right] ~~.
\end{equation}
So, to first order in the interaction, the $\pi$ and $\sigma$ form 
a light multiplet, and the $\eta'$ and $\delta$ a heavy multiplet.
The degeneracy is due to $SU(2)$ chiral symmetry, and the splitting
between the multiplets is a manifestation of $U(1)_A$ violation. Of
course, because the interaction in the scalar channel is so strong, 
the vacuum is rearranged and chiral symmetry is broken spontaneously.
The pion becomes a Goldstone boson and is exactly massless in the 
chiral limit. The sigma corresponds to the massive excitation of 
the quark condensate and is pushed up to $\sim 500-600$ MeV. But
it is clear that the interaction at short distances remains 
attractive, so the sigma will always be lighter than the delta. 

 Short distance perturbative interactions cannot account for 
OZI-violation in the scalar channels. The reason is that 
chirality is strictly conserved in perturbative QCD. The only possible
annihilation channels are $\bar q_Lq_L$ or  $\bar q_Rq_R$,
which do not contribute to the scalar or pseudoscalar channel.
Perturbative effects do contribute to OZI violation in the vector
channel. The OZI violating amplitudes in the vector channel 
(which are suppressed by more than an order of magnitude)
then provide an estimate of the relative role of instanton/pQCD 
effects.

  Phenomenologically, the situation in the scalar $0^{++}$ channel 
appears confused, because of the difficulties in classifying the 
observed scalar mesons. But what is important here is not whether 
some state is too broad to be considered a ``true'' resonance, 
or mixes strongly with $\pi\pi$ or $\bar{K}K$, etc. In fact it is 
pretty clear that there is strong evidence for substantial strength 
in the $\sigma$ channel at energies below 800 MeV. This strength is 
seen experimentally as the strong rise in the $I=0$ $\pi\pi$ phase 
shift and the ``$\sigma$" meson of nuclear physics.

\section{How to measure the OZI-violating amplitude}

 In order to measure the OZI-violating amplitudes we have to 
replace the mass matrices introduced above by objects more directly 
related to field theory. In particular, we shall consider correlation 
functions involving the scalar and pseudoscalar currents introduced 
above. The correlators are defined by
\be
\Pi_\pi(x,y) &=& \langle {\rm Tr}\left(
 S(x,y)i\gamma_5 S(y,x)i\gamma_5\right)\rangle, \\
\Pi_\delta(x,y) &=& \langle {\rm Tr}\left(
 S(x,y) S(y,x)\right)\rangle ,\\
\Pi_{\eta'}(x,y) &=& \langle {\rm Tr}\left(
 S(x,y)i\gamma_5 S(y,x)i\gamma_5\right)\rangle \nonumber \\
 & & \hspace{0.5cm}\mbox{}
 - 2 \langle {\rm Tr}\left(i\gamma_5 S(x,x)\right)
           {\rm Tr}\left(i\gamma_5 S(y,y)\right)\rangle ,\\
\Pi_{\sigma}(x,y) &=& \langle {\rm Tr}\left(
 S(x,y) S(y,x)\right)\rangle \nonumber \\
 & & \hspace{0.5cm}\mbox{}
 - 2 \langle {\rm Tr}\left(S(x,x)\right)
           {\rm Tr}\left(S(y,y)\right)\rangle ,
\ee
where $S(x,y)$ is the fermion propagator and $\langle.\rangle$ 
denotes the average over all gauge configurations. The OZI-violating
difference between the $\pi,\eta'$ and $\sigma,\delta$ channels is
determined by ``disconnected'' (or double hairpin) contributions 
to the correlation functions. Note that it is important to 
correctly define the gamma matrices in the correlation functions
in order to ensure positivity and the existence of a spectral 
representation. In particular, we have to use $\Gamma=i\gamma_5$
in the pseudoscalar channel in order to guarantee $\Pi_\pi(x,y)
>0$. If this is taken care of, we can directly determine the 
sign of the OZI-violating amplitude from the sign of the 
disconnected correlation function. There is a subtlety in the 
scalar $\sigma$ channel, because we need to subtract the 
constant $\langle\bar\psi\psi\rangle^2$ term from the correlation
function. 

\begin{figure}
\begin{center}
\leavevmode
\epsfxsize=7cm
\epsffile{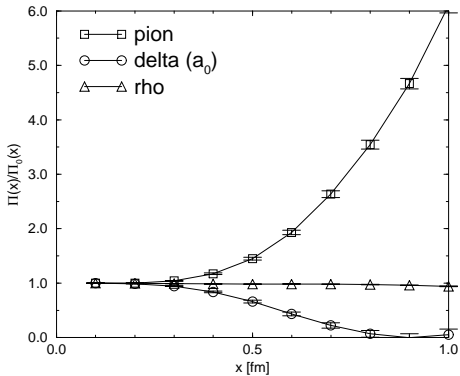}
\end{center}
\caption{Correlation functions in the pion, delta ($a_0$), 
and rho meson channels. The correlators are normalized 
to free field behavior ($\Pi_0(x)\sim x^{-6}$).}
\end{figure}

 Isgur and Thacker find, in the revised version of their 
paper, that the disconnected correlation function is
large and negative in the pseudoscalar channel, large 
and positive in the scalar channel, and very small 
in the vector channel. In Figs. 1 and 2 we compare 
these results to correlation functions obtained in
unquenched instanton simulations \cite{SS_96,SS}. The
correlation functions clearly show {\em exactly the same 
pattern}. There are some technical differences as compared 
to the work of Isgur and Thacker. We measure point-to-point
rather than point-to-plane correlators, and the 
simulations are unquenched rather than quenched. 
This means that there is no need to extract amputated
matrix elements, one can just measure the masses 
directly. We find a heavy $\eta'$ and $\delta$, 
$m_\delta\simeq m_{\eta'}\geq 1$ GeV, and a light
sigma meson, $m_\sigma\simeq 600$ MeV. 

\begin{figure}
\begin{center}
\leavevmode
\epsfxsize=7cm
\epsffile{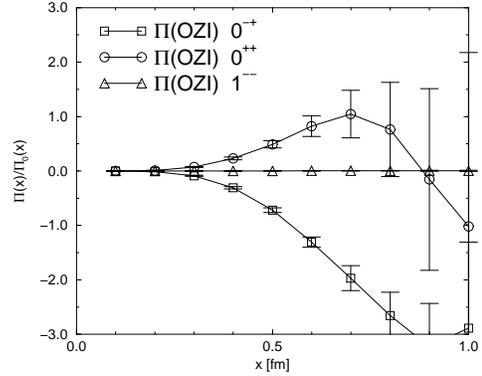}
\end{center}
\caption{OZI violating disconnected correlation functions 
in the pseudoscalar ($\eta'-\pi$), scalar ($\sigma-\delta$),
and vector ($\omega-\rho$) channel. The correlators are 
normalized as in Fig.1. The scalar correlation function 
is very noisy for large $x$ because of the need to make
a subtraction.}
\end{figure}

\section{The large $N_c$ limit}

 The basis of the large $N_c$ approach is the assumption that
$N_c=3$ QCD is similar to QCD in the limit $N_c=\infty$. In
particular, it is assumed that there are no phase transitions
as we go from $N_c=3$ to $N_c\to\infty$. Currently, the status
of these assumptions is not clear, because not much is known
about QCD($N_c=\infty$).

 Let us start with what is firmly known. In the large $N_c$ limit
gauge invariant quantities do not fluctuate. This means that all
physical quantities can be obtained from a classical master 
field. For many years, there was no progress in obtaining the
master field from QCD, except in the case of zero dimensions,
or $SU(N)$ matrix models. Recently, Maldacena discovered a 
master field for $N=4$ supersymmetric QCD. This master field
was described as a certain gravitational metric, together with 
a set of rules that relate gauge theory to supergravity observables.
Amazingly, the same correlation functions can be obtained from
an {\em instanton master field} \cite{Mattis}. This configuration
is a coherent superposition of many instantons with different colors 
orientations but the same size and position, held together by fermion 
exchanges. Progress was also made in understanding $N=2$ SUSY QCD.
Seiberg and Witten determined the low energy effective action of 
this theory. In the semi-classical limit, their result can be 
expressed as a one-loop contribution plus an infinite series 
on $n$-instanton corrections, and {\em nothing else}. Witten
and Seiberg's result was generalized to arbitrary $N_c$ by 
Douglas and Shenker \cite{DS_95}. They identified a special 
form of the large $N_c$ limit in which instantons (and monopoles)
survive. 

  In practice the large $N_c$ expansion is used to determine
the relative importance of certain classes of diagrams. In order 
to have a sensible large $N_c$ limit one requires $g^2N_c={\rm 
const}$. We emphasize that the results are based on the usual
perturbative expansion around the trivial vacuum. Instantons, 
and other non-perturbative effects, can spoil large $N_c$ 
counting rules. Applying large $N_c$ rules to hadronic
correlation functions leads to the usual predictions
\be
M_{generic\,meson} &\sim& N_c^0 M, \\
M_{generic\,baryon}&\sim& N_c^1 M, \\
M_{\eta'}&\sim& N_c^{-1} M,
\label{masses}
\ee
where $M$ is some mass scale of order $\Lambda_{QCD}$. This
prediction is a little bit of an embarrassment, because in the 
real world the nucleon and the $\eta'$ are very close in mass. 
Of course, one can argue that the numerical coefficient in the 
different channels could be very different. Nevertheless, the 
large mass scale that appears in the $\eta'$ channel shows 
that large $N_c$ rules do not work equally well in all 
channels. This observation repeats itself in the scalar 
$0^{++}$ channel, where deviations from large $N_c$ counting
are unusually charge. On the other hand, large $N_c$ 
arguments have proved to be useful in analyzing many
properties of octet and decuplet baryons.

 So what is the $N_c$ dependence of the instanton effects? In his
well known paper \cite{Wit_79}, Witten argued that one has to choose
between instantons and the large $N_c$ explanations of the $U(1)_A$ 
problem, because instanton effects scale as $\exp(-N_c)$. However, 
twenty years later the dilemma does not seem that clear cut. New
arguments have appeared (many by Witten himself), and instantons
and the large $N_c$ limit may well be reconciled one day.  

(i) The suppression of instantons in the large $N_c$ limit
seems to follow from the instanton amplitude $\exp(-8\pi^2/g^2)$, 
together with the 't Hooft scaling $g^2\sim 1/N_c$. But this applies 
to small instantons only. For large instantons $\rho\sim\Lambda^{-1}$ 
the action is $O(1)$, and there is no suppression. This is the 
scenario that was suggested for the $CP^N$ model: For small $N$ 
instantons are small and semi-classical, but for large 
$N$ instantons become strongly overlapping. The question
of the instanton size distribution in QCD is a very non-trivial 
dynamical question. From lattice calculations we only know that the
typical instanton size is about the same in $N_c=2$ and 
$N_c=3$ QCD, $\rho\simeq 1/3$ fm.

(ii) Collecting all factors of $N_c$ in the one-loop instanton
amplitude one finds that they all exponentiate, giving $dN/d\rho 
\sim \exp(N_c F(\rho))$ \cite{SS}. The function $F(\rho)$ has a
non-trivial zero, so the instanton density may scale like a 
power of $N_c$, rather than an exponential, provided the 
instanton distribution becomes a delta function $\delta(\rho-\rho_0)$ 
where $\rho_0$ is the zero of $F(\rho)$.

(iii) The second statement in Witten's paper is that, even in the
absence of instantons, the $\eta'$ mass can be related to the
topological susceptibility $\chi_{top}$ measured in {\em pure gauge} 
theory
\be 
\frac{f_\pi^2}{2N_f} (m_\eta^2+m_{\eta'}^2-2m_K^2) = \chi_{top}.
\ee
This prediction has been checked on the lattice and it works quite
well. After much effort, the lattice measurements of $\chi_{top}$
are stable, and it has become clear that the topological susceptibility
is dominated by instantons, and not by some mysterious fluctuation. 

(iv) As we emphasized above the large $N_c$ limit of both $N=2$ 
and $N=4$ SUSY QCD is consistent with the survival of instantons
in this limit. The $N=2$ case is particularly interesting, because
there are two different ways to take the large $N_c$ limit. In the
``naive" limit, $W$ bosons have masses of $O(1)$, monopoles have
masses $O(N_c)$, and instantons have action $O(N_c)$ and are not
important. In this case, the one loop perturbative result is exact
and nothing interesting happens. A different large $N_c$ limit is 
possible near the singular points on the space of vacua where 
monopoles condense. In this regime one has to tune the Higgs VEVs 
such that $|v_i-v_j| \sim 1/N_c$. Then monopole masses are $O(1)$, 
and instantons are not suppressed. In this case the naive scaling
relation $g^2\sim 1/N_c$ is not satisfied.

(v) The large $N_c$ prediction for pure gluodynamics $\chi_{top} 
\sim \Lambda_{QCD}^4 \sim N_c^{-2} \epsilon_{vac}$ was recently
verified in a string theory setting \cite{Wit_98}. This result 
is consistent with a number of scenarios for the instanton liquid.
For example, the instanton ensemble might evolve from a random
gas of density $n\sim \chi_{top}$ at small $N_c$ to a strongly 
correlated liquid with large density $(N/V)\sim\epsilon_{vac}$ but
very small fluctuations $\chi_{top}=\langle(N_I-N_A)^2\rangle
/V\sim N_c^{-2}(N/V)$.

(vi) Instantons are not incompatible with the success of large 
$N_c$ arguments in the baryon sector. The instanton induced 
interaction between quarks is $1/N_c$ suppressed, just like 
one gluon exchange. Also, the topological soliton model can be 
derived from instantons in the large $N_c$ limit \cite{Dia_97}.
This model automatically incorporates all the large $N_c$ 
predictions.

\section{Smallness of OZI-violating amplitudes in channels
other than $O^\pm$: conspiracy between hadronic amplitudes}

  We completely agree with Isgur and Thacker on this point. We
would like to make two comments. First of all, instantons are 
compatible with the smallness of the OZI violating amplitude in
the vector channel. The direct instanton contribution to the 
disconnected correlator vanishes. The next correction is 
suppressed also. Correlated instanton-anti-instanton pairs
contribute to connected, but not to disconnected vector
correlators \cite{SSV_95}.
 
 Second, we would like to point out that we have compiled
an extensive phenomenological analysis of hadronic correlation 
functions \cite{Shu_cor}. This compilation contains a number of 
examples for cancellations between a large number of hadronic
amplitudes. An example is the striking ``superduality'' 
phenomenon in the vector channels\footnote{We note that 
this phenomenon is different from the standard parton-hadron
duality. Ordinary duality holds in all channels, but usually
breaks down at distances $x>0.3$ fm.}: The full correlation
function remains close to free field behavior for
distances as large as $|x-y|<1.3$ fm. In terms of the
spectral representation, this behavior requires remarkable
fine tuning between different hadronic amplitudes. This 
result can be directly verified from data, using the 
cross section for $e^+e^-$ annihilation into hadrons.

  Another important conclusions from this study is the fact 
that OZI violation in the scalar and pseudoscalar channel
appears at very short distances. This is one more argument 
in favor of instantons rather than confinement as the source
of OZI violation. While instantons appear at short distances,
confinement is a long distance effect.

\section{Conclusions}

 In summary we showed that instantons explain the sign and magnitude
of the OZI violating amplitude in all mesonic channels, not just 
for the $\eta'$. In addition to that, instantons provide a 
mechanism for chiral symmetry breaking, and a successful 
description of light hadron spectroscopy. The role of instantons 
in the large $N_c$ limit is a complicated issue, but instantons 
and the large $N_c$ need not be incompatible. Nevertheless, naive 
large $N_c$ counting rules may not work in all channels. Instantons 
provide an explanation why large $N_c$ predictions work well in some 
cases, but fail badly in others.

\end{narrowtext}
\end{document}